\documentclass[aps,prc,superscriptaddress,floats,floatfix,reprint,reprintnumbers]{revtex4}
\usepackage{blindtext}
\pdfoutput=1
\usepackage{epsfig}
\usepackage{graphicx}% Include figure files
\usepackage{dcolumn}% Align table columns on decimal point
\usepackage{bm}% bold math
\usepackage{booktabs}
\usepackage{color}
\usepackage{mathrsfs}
\usepackage{ulem}
\usepackage{epstopdf}
\graphicspath{{plots/}}%directory of my plots
\usepackage{verbatim}
\usepackage{indentfirst}
\usepackage{float}
\usepackage[colorlinks,
           bookmarks=true,
           linkcolor=blue,
           urlcolor=blue,
            anchorcolor=black,
            citecolor=blue]{hyperref}
\usepackage[bf,raggedright,indentafter]{titlesec}
\usepackage{rotating}
\usepackage{booktabs}
\usepackage{tabularx}
\usepackage{makecell}
\usepackage{hypcap}
\usepackage{subfigure}%

\renewcommand{\baselinestretch}{1.47}

\begin{document}
\title{\Large Studying Two-body Nonleptonic  Weak Decays of Hyperons with Topological Diagram Approach  }
\author{Yuan-Guo Xu$^{1,\dag}$, ~Xiao-Dong Cheng$^{2,\ddag}$, ~Jie-Lei Zhang$^{2,\sharp}$ ~and ~Ru-Min Wang$^{1,\S}$\\
$^1${\scriptsize College of Physics and Communication Electronics, JiangXi Normal University, NanChang, JiangXi 330022, China}\\
$^2${\scriptsize College of Physics and Electronic Engineering, XinYang Normal University, XinYang, Henan 464000, China}\\
$^\dag${\scriptsize yuanguoxv@163.com}~~~~~~~
$^\ddag${\scriptsize chengxd@mails.ccnu.edu.cn}\\
$^\sharp${\scriptsize zhangjielei@ihep.ac.cn}~~~~~~
$^\S${\scriptsize ruminwang@sina.com~~~~~~}\\
}

\begin{abstract}
Many decays of light baryons consisting of light $u,d,s$ quarks have been measured,  and these measurements will help us to understand the decay properties of light baryons.  In this work, we study two-body nonleptonic weak decays of light baryon octet $(T_8)$ and baryon decuplet $(T_{10})$ by the topological diagram approach (TDA) under the SU(3) flavor
symmetry for the first time. We find that (1) the TDA and the SU(3) irreducible representation approach (IRA) match consistently in $T_{10}\to T^{(')}_{8,10}P_8$ ($P_8$ is the light pseudoscalar meson octet); (2)    almost all  relevant not-yet-measured $\mathcal{B}(T_{10}\to T_8 P_8)$ may be predicted by using three experimental data of $\mathcal{B}(\Omega^-\to \Xi^0\pi^-,\Xi^-\pi^0,\Lambda^0K^-)$, and the upper limits of $\mathcal{B}(T_{10}\to T{'}_{10}\pi^-)$ may be obtained from the experimental upper limit of $\mathcal{B}(\Omega^-\to \Xi^{*0}\pi^-)$ by both the TDA and the IRA together, nevertheless, all new predicted branching ratios  are too small to br reached in current experiments; (3) $T_8\to T'_8 P_8$ decays  are quite complex in terms of the TDA,  and we find that W-exchange diagrams give large and even dominant contributions by using relevant experimental data and  the isospin relations.

\end{abstract}

\maketitle

\section{INTRODUCTION}

Many two-body nonleptonic weak decays of the octet and decuplet baryons   (such as $\Sigma^+\rightarrow p\pi^0$,
$\Sigma^+\rightarrow n\pi^+$, $\Sigma^-\rightarrow n\pi^-$, $\Lambda^0\rightarrow p\pi^-$, $\Lambda^0\rightarrow n\pi^0$, $\Xi^-\rightarrow \Lambda^0\pi^-$, $\Xi^0\rightarrow \Lambda^0\pi^0$, $\Omega^-\rightarrow \Xi^0\pi^-,\Xi^-\pi^0,\Lambda^0K^-$) were measured  a long time ago by SPEC, HBC, OSPK etc \cite{PDG2018}. Now the sensitivity for measurements of $\Lambda,\Sigma,\Xi,\Omega$ hyperon decays  is in the range of $10^{-5}-10^{-8}$ at the BESIII \cite{Li:2016tlt,Bigi:2017eni,Asner:2008nq,BESIIILi2019}, and these hyperons are also produced copiously  at the LHCb experiment~\cite{Aaij:2017ddf,Junior:2018odx}.
The precise measurements from  the  BESIII  and LHCb experiments will confirm the
 earlier experimental data from SPEC, HBC, OSPK etc, and will give  new information about determining the V-A structure and quark-flavor mixing \cite{Weinberg:2009zz,Severijns:2006dr,Cabibbo:1963yz} as well as probing  the non-standard charged current interactions \cite{Cirigliano:2012ab,Chang:2014iba}.

In the theoretical side, the heavy quark expansion can not be used in the light baryon decays, and the factorization does not work well. There is no reliable method to calculate these decay matrix elements at present. In the lack of reliable calculations, some model-independent approaches can provide very useful information about the decays, such as SU(3)/U(3) flavor symmetry (see for instance Refs. \cite{Altarelli:1975ye,Savage:1989qr,Savage:1991wu,Hsiao:2015iiu,He:2015fsa,He:2015fwa,Gronau:2013mza,Arora:1992yq,Du:1994qt,Korner:1994nh,Hsiao:2014mua,Gronau:2015jgh,Egolf:2002nk,Lu:2016ogy,Geng:2018upx,Wang:2019alu,Dery:2020lbc}) and flavor topological diagram approach (TDA) (see for instance Refs. \cite{Zeppenfeld:1980ex,Chau:1986du,Chau:1987tk,Savage:1989ub,Chau:1990ay,Gronau:1994rj,Gronau:1995hn,Wang:2017hxe,Zhou:2016jkv,Cheng:2014rfa,Chiang:2006ih,Muller:2015lua,Zhou:2015jba,Chiang:2008zb,He:2018php}). These approaches are independent of the detailed dynamics, and offer an opportunity to relate different decay modes.  The decay matrix elements are directly  extracted from the experimental data, despite of their unclear sources. In the TDA, decay amplitudes are represented by connecting quark line flows in different ways  and then relate them by the SU(3) symmetry,
therefore, the TDA gives a better understanding of dynamics in the different amplitudes.

The related decays have also been extensively studied,  for instance, in Refs. \cite{He:2019xxp,Flores-Mendieta:2019lao,Jiang:2008aqa,Tandean:2004mv,Borasoy:2003rc,Tandean:2002vy,Scadron:2000wa,AbdElHady:1999mj,Springer:1999sv,Tandean:1998wr,Borasoy:1998ku,He:1997bs,Jenkins:1991bt}.
 The studies in Refs. \cite{Cabibbo:1963yz,Gaillard:1984ny,Carrillo-Serrano:2014zta,Pham:2012db,FloresMendieta:1998ii,Wang:2019alu} show that the SU(3) flavor
symmetry breaking effects  are small in the semileptonic hyperon
decays and the nonleptonic hyperon two-body weak decays.
So, in this paper,  we will  study $T_{10}\to T_{8}M_8$, $T_{10}\to T'_{10}M_8$  and $T_{8}\to T'_{8}M_8$  nonleptonic two-body  weak decays by using the TDA under the SU(3) flavor
symmetry, which is the following up work of Ref. \cite{Wang:2019alu}.
We  will  firstly construct the
TDA amplitudes for different kinds of $T_{8}$ and $T_{10}$ nonleptonic decays,  secondly  obtain the decay amplitude relations between different decay modes, then  extract the TDA amplitudes from  the available data,  and finally  analyze the size of the different kinds of contributions  or predict the  not-yet-measured modes  for further tests in experiments.

This paper is organized as follows. In Sec. II, we give the effective Hamiltonian for nonleptonic $s\to u\bar{u}d$ process and the expression of the branching ratios.
In Sec. III, we firstly give  the TDA amplitudes and then give our numerical results and analyses for $T_{10}\to T_{8}M_8$, $T_{10}\to T'_{10}M_8$  and $T_{8}\to T'_{8}M_8$  decays.  Section IV contains our summary and conclusion.

\section{Theoretical framework}
Under the $SU(3)$ flavor symmetry of $u,d,s$ quarks, the light baryon octet  $T_{8}$, the light baryon decuplet $T_{10}$, the
light pseudoscalar meson octet $P_8$, and the vector meson octet $V_8$ can be written as
\begin{eqnarray}
 T_8&=&\left(\begin{array}{ccc}
\frac{\Lambda^0}{\sqrt{6}}+\frac{\Sigma^0}{\sqrt{2}} & \Sigma^+ & p \\
\Sigma^- &\frac{\Lambda^0}{\sqrt{6}}-\frac{\Sigma^0}{\sqrt{2}}  & n \\
\Xi^- & \Xi^0 &-\frac{2\Lambda^0}{\sqrt{6}}
\end{array}\right)\,,\\
%\end{eqnarray}
%\begin{eqnarray}
 T_{10}&=&\frac{1}{\sqrt{3}}\left(
 \left(\begin{array}{ccc}
\sqrt{3}\Delta^{++} & \Delta^{+}  & \Sigma^{*+}  \\
\Delta^{+} & \Delta^{0}  & \frac{\Sigma^{*0}}{\sqrt{2}}  \\
\Sigma^{*+} & \frac{\Sigma^{*0}}{\sqrt{2}}  &\Xi^{*0}
\end{array}\right),
\left(\begin{array}{ccc}
\Delta^{+} & \Delta^{0}  & \frac{\Sigma^{*0}}{\sqrt{2}}  \\
\Delta^{0} & \sqrt{3}\Delta^{-}  & \Sigma^{*-} \\
\frac{\Sigma^{*0}}{\sqrt{2}} & \Sigma^{*-} &\Xi^{*-}
\end{array}\right),
\left(\begin{array}{ccc}
\Sigma^{*+} & \frac{\Sigma^{*0}}{\sqrt{2}}  &\Xi^{*0}  \\
\frac{\Sigma^{*0}}{\sqrt{2}} & \Sigma^*{-}  & \Xi^{*-} \\
\Xi^{*-} & \Xi^{*-} &\sqrt{3}\Omega^{-}
\end{array}\right)\right),\\
%\end{eqnarray}
%\begin{eqnarray}
 P_8&=&\left(\begin{array}{ccc}
\frac{\eta_8}{\sqrt{6}}+\frac{\pi^0}{\sqrt{2}} & \pi^+ & K^+ \\
\pi^- &\frac{\eta_8}{\sqrt{6}}-\frac{\pi^0}{\sqrt{2}}  & K^0 \\
K^- & \bar{K}^0 &-\sqrt{\frac{2}{3}}\eta_8
\end{array}\right)\,,\\
V_8&=&\left(\begin{array}{ccc}
\frac{\omega_8}{\sqrt{6}}+\frac{\rho^0}{\sqrt{2}} & \rho^+ & K^{*+} \\
\rho^- &\frac{\omega_8}{\sqrt{6}}-\frac{\rho^0}{\sqrt{2}}  & K^{*0} \\
K^{*-} & \bar{K}^{*0} &-\sqrt{\frac{2}{3}}\omega_8
\end{array}\right)\,.
\end{eqnarray}

Two-body nonleptonic weak decays $T_{8,10}\to T'_{8,10}M_8$ with $M_8=P_8,V_8$  are induced by $s\to u\bar{u}d$ transition. As given in Ref. \cite{Wang:2019alu}, there are two kinds of diagrams for s-quark weak decay   in the Standard Model,  the tree level diagram  and the penguin diagram.  The effective Hamiltonian for nonleptonic $s\to u\bar{u}d$  process at scales $\mu<m_c$ can be written as \cite{Buchalla:1995vs}
% 9512380 P67 Eq(VII.1), P53 Eq(VI.3)
\begin{eqnarray}
\mathcal{H}_{eff}=\frac{G_F}{\sqrt{2}}V_{ud}V_{us}^*\sum_{i=1}^{10}\Biggl[z_i(\mu)-\frac{V_{td}V_{ts}^*}{V_{ud}V_{us}^*}y_i(\mu)\Biggl]Q_i(\mu),\label{EQ:Heff}
\end{eqnarray}
where  $V_{uq}$ is the CKM matrix element,  $z_i(\mu)$ and $y_i(\mu)$ are  Wilson coefficients, and  $Q_i$ is  four-quark operators.
The detail can be found in Refs. \cite{Wang:2019alu,Buchalla:1995vs}.

\begin{figure}[b]
\centering
\includegraphics[scale=1]{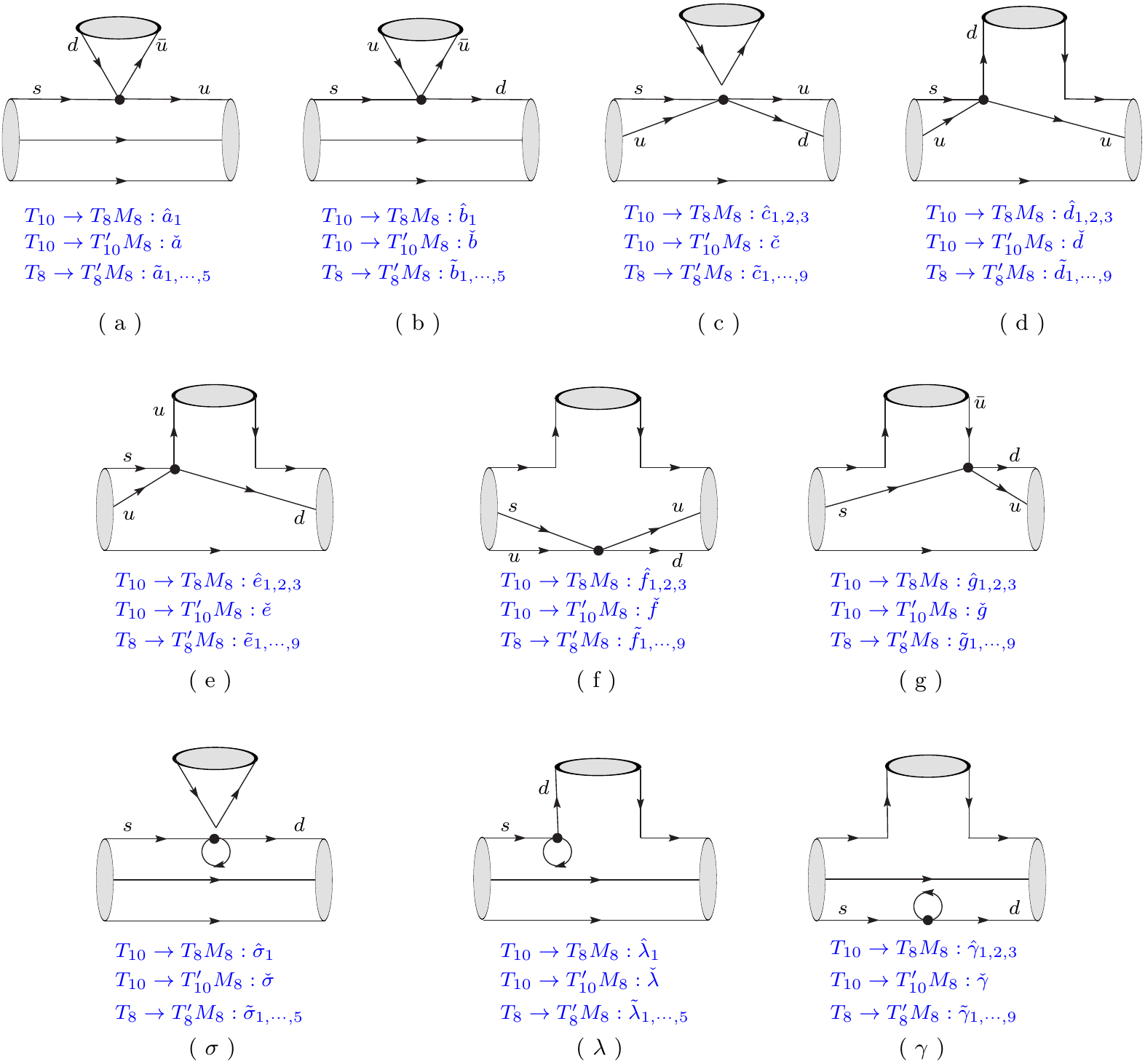}
\caption{Topological diagrams for $T_{8,10}\to T'_{8,10}M_8$ nonleptonic decays. Since the octed baryon $T_{8}$ is not fully asymmetric or antisymmetric in flavor space, there might be more than one amplitudes corresponding
to one relevant topological diagram. }
\label{fig:NLTP}
\end{figure}

The relevant topological diagrams for $T_{8,10}\to T_{8}M_8$ nonleptonic weak decays are displayed in Fig. \ref{fig:NLTP}.
Since the octet baryon is
not fully symmetric or antisymmetric in flavor space, there are more than one amplitudes corresponding to one topological
diagram.
Noted that Fig. \ref{fig:NLTP} (c,$\sigma$) only contributes the decays with $M_8=\eta,\omega$, which will not be considered in this work.  Other topological diagrams in Fig. \ref{fig:NLTP} can be divided into three categories: the tree diagrams in Fig. \ref{fig:NLTP} (a,b), the W-exchange diagrams in Fig. \ref{fig:NLTP} (d,e,f,g)  and the penguin diagrams in Fig. \ref{fig:NLTP} ($\lambda,\gamma$).
Fig. \ref{fig:NLTP} (a) contributes to the decays with the charged $M_8$, which  is proportional to the color-flavor factor $C_2+\frac{C_1}{N_C}$. Fig. \ref{fig:NLTP} (b) contributes to the decays with the neutral $M_8$, which  is proportional to $C_1+\frac{C_2}{N_C}$.  The contributions of Fig. \ref{fig:NLTP} (d,e,f,g) are all  proportional to $C_2-C_1$. Compared with the tree diagrams and W-exchange contributions  related to $C_{1,2}$, the penguin contributions displayed in Fig. \ref{fig:NLTP} ($\lambda,\gamma$) are strongly suppressed by small Wilson coefficients $C_{3,\cdots,10}$ and quite small CKM factor,  and their contributions can be safely ignored in these decays if one do not consider the information of CP violation.

The branching ratios of the $T_A\to T_B M_{C}$ two-body decays can be written  in terms of the decay amplitudes $A(T_{A}\to T_{B} M_C)$
\begin{eqnarray}
\mathcal{B}(T_{A}\to T_{B} M_C)=\frac{\tau_A|p_{cm}|}{8\pi m_A^2}\big|A(T_{A}\to T_{B} M_C)\big|^2S,\label{Eq:Br}
\end{eqnarray}
where $T_{A,B}=T_{8}$ or $T_{10}$, $M_C=M_8$,  $|p_{cm}|=\sqrt{(m_A^2-(m_B+m_C)^2)(m_A^2-(m_B-m_C)^2)}/(2m_A)$, $S=1$ for $T_A=T_8$, and $S=1/2$ for $T_A=T_{10}$.  For more accurate results, we will consider the mass difference in $A(T_{A}\to T_{B} P_8)$  from  Eqs. (66-68) of Ref. \cite{He:2018joe}
\begin{eqnarray}
A(T_{A}\to T_{B} P_8)\propto\frac{m_A}{m_B}p_{cm}N_BN_A,\label{Eq:AMDT8}
\end{eqnarray}
with $N_A=\sqrt{2m_A}$ and $N_B=\sqrt{((m_A+m_B)^2-m_P^2)/(2m_A)}$.  The amplitudes  $A(T_{A}\to T_{B} M_C)$ for different decay processes are given in next section.

\section{Results and discussions}

In this section, we only give the concrete amplitudes of  $T_{8,10}\to T'_{8,10}P_8$ decays,  and the corresponding amplitudes of  $T_{8,10}\to T'_{8,10}V_8$ decays can be obtained similarly by replacing $P_8$ to $V_8$  in following Tab. \ref{Tab:T102T8PTDA}, Tab. \ref{Tab:T102T10PTDA} and Tab. \ref{Tab:T82T8PTDA}.
For $T_8\to T_{10}M_8$ decays with $s\to u\bar{u}d$ transition, we find that these  processes are not allowed by the energy conservation law.   So we will only give the  detail analyses for $T_{10}\to T_8P_8$, $T_{10}\to T'_{10}P_8$ and $T_8\to T'_8P_8$ weak decays. In addition, the theoretical input parameters and the experimental data within the $1\sigma$ error from Particle Data Group \cite{PDG2018} will
be used in our numerical results.

\subsection{$T_{10}\to T_8M_8$ weak decays}

Topological diagrams for $T_{10}\to T_{8}M_8$ weak decays are displayed in Fig. \ref{fig:NLTP}, and the TDA amplitudes of the $T_{10}\to T_8M_8$ weak decays can be written as
{\small
\begin{eqnarray}
A(T_{10}\to T_8M_8)^{TDA}&=& \hat{a}_1H'^{kl}_{mn}(T_{10})^{nij}(T_{8})_{[ik]j}(M_8)_l^m\nonumber\\
&+&\hat{b}_1H'^{kl}_{mn}(T_{10})^{nij}(T_{8})_{[il]j} (M_8)_k^m\nonumber\\
&+&\hat{c}_1H'^{il}_{kn}(T_{10})^{nij}(T_{8})_{[jk]l}(M_8)_m^m+\hat{c}_2H'^{il}_{kn}(T_{10})^{nij}(T_{8})_{[jl]k} (M_8)_m^m+\hat{c}_3H'^{il}_{kn}(T_{10})^{nij}(T_{8})_{[kl]j} (M_8)_m^m\nonumber\\
&+&\hat{d}_1H'^{lk}_{in}(T_{10})^{nij}(T_{8})_{[jm]k}(M_8)_l^m+\hat{d}_2H'^{lk}_{in}(T_{10})^{nij}(T_{8})_{[jk]m} (M_8)_l^m+\hat{d}_3H'^{lk}_{in}(T_{10})^{nij}(T_{8})_{[km]j} (M_8)_l^m\nonumber\\
&+&\hat{e}_1H'^{kl}_{in}(T_{10})^{nij}(T_{8})_{[jm]l}(M_8)_k^m+\hat{e}_2H'^{kl}_{in}(T_{10})^{nij}(T_{8})_{[jl]m} (M_8)_k^m+\hat{e}_3H'^{kl}_{in}(T_{10})^{nij}(T_{8})_{[lm]j} (M_8)_k^m\nonumber\\
&+&\hat{f}_1H'^{kl}_{in}(T_{10})^{nij}(T_{8})_{[kl]m}(M_8)_j^m+\hat{f}_2H'^{kl}_{in}(T_{10})^{nij}(T_{8})_{[km]l} (M_8)_j^m+\hat{f}_3H'^{kl}_{in}(T_{10})^{nij}(T_{8})_{[ml]k} (M_8)_j^m\nonumber\\
&+&\hat{g}_1H'^{lk}_{mn}(T_{10})^{nij}(T_{8})_{[ik]l}(M_8)_j^m+\hat{g}_2H'^{lk}_{mn}(T_{10})^{nij}(T_{8})_{[il]k} (M_8)_j^m+\hat{g}_3H'^{lk}_{mn}(T_{10})^{nij}(T_{8})_{[lk]i} (M_8)_j^m\nonumber\\
&+&\hat{\sigma}_1H'^{lk}_{ln}(T_{10})^{nij}(T_{8})_{[ik]j} (M_8)_m^m\nonumber\\
&+&\hat{\lambda}_1H'^{lk}_{ln}(T_{10})^{nij}(T_{8})_{[im]j} (M_8)_k^m\nonumber\\
&+&\hat{\gamma}_1H'^{lk}_{ln}(T_{10})^{nij}(T_{8})_{[ik]m}(M_8)_j^m+\hat{\gamma}_2H'^{lk}_{ln}(T_{10})^{nij}(T_{8})_{[im]k} (M_8)_j^m+\hat{\gamma}_3H'^{lk}_{ln}(T_{10})^{nij}(T_{8})_{[mk]i} (M_8)_j^m,~~~
\label{T102T8M8TDA}
\end{eqnarray}}
where $H'^{kl}_{mn}\equiv V_{mn}V^*_{kl}$ containing the CKM factors.

The TDA amplitudes for the $T_{10}\to T_8 P$ decays are given in Tab. \ref{Tab:T102T8PTDA}.
\begin{table}[b]
\renewcommand\arraystretch{1.38}
\tabcolsep 0.15in
\centering
\caption{The TDA amplitudes of the $T_{10}\to T_8 P$ decays under the $s\to u\bar{u}d$ transition. Noted that $\hat{X}_1\equiv\tilde{x}_1-\tilde{x}_2+2\hat{x}_3$ and $\hat{X}_2\equiv\tilde{x}_1+\tilde{x}_2$ with $\hat{X}=\hat{G},\hat{\Gamma},\hat{D},\hat{E}$ for $\hat{g}_i,\hat{\gamma}_i,\hat{d}_i,\hat{e}_i$, respectively. }\vspace{0.1cm}
{\footnotesize
\begin{tabular}{r|ccc}  \hline\hline
 Observables & Tree& W-exchange&Penguin\\\hline
$A(\Omega^-\rightarrow\Xi^0\pi^-)$&$\hat{a}_1$&&$+\hat{\beta}_1$ \\
$\sqrt{2}A(\Omega^-\rightarrow\Xi^-\pi^0)$&$-\hat{b}_1$&&$+\hat{\beta}_1$ \\
$\sqrt{6}A(\Omega^-\rightarrow\Lambda^0K^-)$&&$-\hat{G}_1$&$-\hat{\Gamma}_1$ \\
\hline
$3\sqrt{2}A(\Xi^{*-}\rightarrow\Lambda^0\pi^-)$&$3\hat{a}_1$&$-\hat{G}_1$&$+3\hat{\beta}_1-\hat{\Gamma}_1$ \\
$\sqrt{6}A(\Xi^{*-}\rightarrow\Sigma^0\pi^-)$&$\hat{a}_1$&$+\hat{G}_2$&$+\hat{\beta}_1+\hat{\Gamma}_2$\\
$\sqrt{6}A(\Xi^{*-}\rightarrow\Sigma^-\pi^0)$&$-\hat{b}_1$&&$+\hat{\beta}_1+\hat{\Gamma}_2$\\
$\sqrt{3}A(\Xi^{*0}\rightarrow\Sigma^+\pi^-)$&$\hat{a}_1$&$+\hat{D}_2$&$+\hat{\beta}_1$\\
$6A(\Xi^{*0}\rightarrow\Lambda^0\pi^0)$&$-3\hat{b}_1$&$+\hat{D}_1+\hat{E}_1-\hat{G}_1$&$+3\hat{\beta}_1-\hat{\Gamma}_1$\\
$2\sqrt{3}A(\Xi^{*0}\rightarrow\Sigma^0\pi^0)$&$\hat{b}_1$&$-\hat{D}_2+\hat{E}_2+\hat{G}_2$&$-\hat{\beta}_1+\hat{\Gamma}_2$\\
$\sqrt{3}A(\Xi^{*0}\rightarrow\Sigma^-\pi^+)$&&$-\hat{E}_2$&$+\hat{\Gamma}_2$\\
$\sqrt{3}A(\Xi^{*-}\rightarrow nK^-)$&&$(\hat{G}_1-\hat{G}_2)/2$&$-(\hat{\Gamma}_2-\hat{\Gamma}_1)/2$\\
\hline
$\sqrt{3}A(\Sigma^{*-}\rightarrow n\pi^-)$&$-\hat{a}_1$&$+(\hat{G}_1-\hat{G}_2)/2$&$-\hat{\beta}_1-(\hat{\Gamma}_2-\hat{\Gamma}_1)/2$\\
$\sqrt{6}A(\Sigma^{*0}\rightarrow p\pi^-)$&$-\hat{a}_1$&$-\hat{D}_2+(\hat{F}_1+\hat{F}_2)/2+(\hat{G}_1+\hat{G}_2)/2$&$-\hat{\beta}_1+(\hat{\Gamma}_1+\hat{\Gamma}_2)/2$\\
$2\sqrt{3}A(\Sigma^{*0}\rightarrow n\pi^0)$&$\hat{b}_1$&$(\hat{D}_2-\hat{D}_1)/2-(\hat{E}_1+\hat{E}_2)/2-\hat{F}_2+(\hat{G}_1-\hat{G}_2)/2$&$-\hat{\beta}_1-(3\hat{\Gamma}_2-\hat{\Gamma}_1)/2$\\
$\sqrt{6}A(\Sigma^{*+}\rightarrow p\pi^0)$&$\hat{b}_1$&$-(\hat{D}_2+\hat{D}_1)/2+(\hat{E}_2-\hat{E}_1)/2+(\hat{F}_1+\hat{F}_2)/2+(\hat{G}_1+\hat{G}_2)/2$&$-\hat{\beta}_1+(\hat{\Gamma}_1+\hat{\Gamma}_2)/2$\\
\hline
\end{tabular}\label{Tab:T102T8PTDA}}
\end{table}
Noted that $\hat{f}_i$ terms in Fig. \ref{fig:NLTP} (f) related to  the $T_{10}\to T_{8}M_8$ decays only contribute to $\Sigma^{*0}\to p \pi^-$, $\Sigma^{*0}\to n \pi^0$ and $\Sigma^{*+}\to p \pi^0$ processes.
Comparing the  SU(3) IRA amplitudes in Tab. VIII of Ref. \cite{Wang:2019alu} with the TDA amplitudes  in  Tab. \ref{Tab:T102T8PTDA}, we have the following relations
\begin{eqnarray}
\bar{a}_1&=&\frac{\hat{a}_1+\hat{b}_1}{4},~~~~~~~~~~~~\bar{e}_1=\frac{\hat{a}_1-\hat{b}_1+2\hat{\beta}_1}{4},~~~~~~~~~~~~\bar{b}_2=-\frac{\hat{G}_2}{4},\nonumber\\
\bar{c}_3+\bar{c}_2&=&\hat{D}_2=\hat{E}_2,~~~~~~~~~~2\bar{c}_1+\bar{c}_2+\bar{c}_3=\frac{\hat{D}_1+\hat{E}_1}{2},~~~~~~~~~~~~2\bar{d}_1=-\hat{F}_2,\nonumber\\
\bar{f}_1+\bar{f}_2&=&\frac{\hat{G}_1+\hat{\Gamma}_1}{2}=-\hat{F}_1,~~~~~~~~~~\bar{f}_1=\bar{f}_3,~~~~~~~~~~~~~~~~~~~~~~~~~~~~\hat{\Gamma}_2=0.
\label{Eq:TDAIRAR}
\end{eqnarray}
From Eq. (\ref{Eq:TDAIRAR}) one can see that the parameters $\hat{G}_2,\hat{D}_1+\hat{E}_1,\hat{D}_2$, $\hat{E}_2$ and  $\hat{F}_2$  terms in TDA amplitudes are related to $\bar{b}_2$, $\bar{c}_{1,2,3}$ or $\bar{d}_1$ terms in IRA amplitudes, which only  appear in the Wilson coefficient suppressed  $C_+$ term, so we will neglect their contributions in following numerical analysis.

Using the amplitude relations listed in Tab. \ref{Tab:T102T8PTDA} and the experimental data of $\mathcal{B}(\Omega\to \Xi^0\pi^-,\Xi^-\pi^0,\Lambda^0K^-)$, we obtain  all  branching ratio predictions for relevant $\Xi^{*-},\Xi^{*0}$ and $\Sigma^{*-}$ decays, which are  listed in the last column of Tab. \ref{Tab:BrT102T8M8TDA}.
Noted that the mass difference in the amplitudes  similar to Eq. (\ref{Eq:AMDT8}) is also considered.
\begin{table}[t]
\renewcommand\arraystretch{1.38}
\tabcolsep 0.30in
\centering
\caption{Branching ratios of the $T_{10}\to T_8P_8$ decays within $1\sigma$ error under the $s\to u\bar{u}d$ transition. $^\dagger$ denotes that the predictions depend on the relative phases. }\vspace{0.1cm}
{\footnotesize
\begin{tabular}{lccc}  \hline\hline
 Branching ratios &Exp. \cite{PDG2018}&TDA \\\hline
$\mathcal{B}(\Omega^-\rightarrow\Xi^0\pi^-)(\times10^{-2})$&$23.6\pm0.7$&$23.6\pm0.7$\\
$\mathcal{B}(\Omega^-\rightarrow\Xi^-\pi^0)(\times10^{-2})$&$8.6\pm0.4$&$8.6\pm0.4$ \\
$\mathcal{B}(\Omega^-\rightarrow\Lambda^0K^-)(\times10^{-2})$&$67.8\pm0.7$&$67.8\pm0.7$  \\
\hline
$\mathcal{B}(\Xi^{*-}\rightarrow\Lambda^0\pi^-)(\times10^{-12})$&$\cdots$&$1.06\pm0.90^\dag$\\
$\mathcal{B}(\Xi^{*-}\rightarrow\Sigma^0\pi^-)(\times10^{-14})$&$\cdots$&$2.87\pm0.65$\\
$\mathcal{B}(\Xi^{*-}\rightarrow\Sigma^-\pi^0)(\times10^{-14})$&$\cdots$&$2.14\pm0.52$\\
$\mathcal{B}(\Xi^{*0}\rightarrow\Sigma^+\pi^-)(\times10^{-14})$&$\cdots$&$5.96\pm0.58$\\
$\mathcal{B}(\Xi^{*0}\rightarrow\Lambda^0\pi^0)(\times10^{-13})$&$\cdots$&$5.02\pm4.06^{\dag}$\\
$\mathcal{B}(\Xi^{*0}\rightarrow\Sigma^0\pi^0)(\times10^{-14})$&$\cdots$&$1.12\pm0.13$\\
$\mathcal{B}(\Xi^{*0}\rightarrow\Sigma^-\pi^+)$&$\cdots$&$0$\\
$\mathcal{B}(\Xi^{*-}\rightarrow nK^-)(\times10^{-13})$&$\cdots$&$6.14\pm1.31$\\
%$\mathcal{B}(\Xi^{*0}\rightarrow pK^-)$&$\cdots$&$$\\
\hline
$\mathcal{B}(\Sigma^{*-}\rightarrow n\pi^-)(\times10^{-13})$&$\cdots$&$3.84\pm2.41^{\dag}$\\
$\mathcal{B}(\Sigma^{*0}\rightarrow p\pi^-)(\times10^{-13})$&$\cdots$&$2.20\pm1.48^{\dag}$\\
$\mathcal{B}(\Sigma^{*0}\rightarrow n\pi^0)(\times10^{-13})$&$\cdots$&$1.07\pm0.68^{\dag}$\\
$\mathcal{B}(\Sigma^{*+}\rightarrow p\pi^0)(\times10^{-13})$&$\cdots$&$2.03\pm1.24^{\dag}$\\
\hline
\end{tabular}\label{Tab:BrT102T8M8TDA}}
\end{table}
In Tab. \ref{Tab:BrT102T8M8TDA}, one can see that $\mathcal{B}(\Omega\to \Xi^{*-}\pi^0,\Xi^{*0}\pi^-,\Sigma^{*0}K^-)$ are on the order of $\mathcal{O}(10^{-2}-10^{-1})$, nevertheless, all branching ratios of $\Xi^{*}$ and $\Sigma^{*}$ weak decays  are  on the order of $\mathcal{O}(10^{-14}-10^{-12})$.   The main reason of significant magnitude difference comes the different decay mechanisms between $\Omega$ decays and $\Xi^{*}(\Sigma^{*})$ decays. $\Omega$ decays  such as $\Omega\to \Xi^{*-}\pi^0,\Xi^{*0}\pi^-,\Sigma^{*0}K^-$ decay modes  mainly through weak interactions,  and its life time $\tau_\Omega\approx0.821\times10^{-10}s$. But $\Xi^{*}$ and $\Sigma^{*}$ decay mainly through strong interactions, such as $\Xi^{*}\to \Xi \pi$ as well as $\Sigma^{*}\to \Lambda \pi,\Sigma\pi$, and their life times are on the order of  $10^{-23}s$.  From Eq. (\ref{Eq:Br}), assuming all decay amplitudes are similar, the  $\Xi^{*}$ and $\Sigma^{*}$ branching ratios are smaller than $\Omega$ decay branching ratios by about $10^{-12}$ orders.   The similar reason for later $T_{10}\to T'_{10}P_8$ decays listed in Tab. \ref{Tab:T102T10PRBr}.

\subsection{$T_{10}\to T'_{10}M_8$  weak decays}
For  $T_{10}\to T'_{10}M_8$ decays, their amplitudes are simple since both initial states $T_{10}$ and final states $T'_{10}$ are all fully symmetric in flavor space. The  TDA amplitudes of the $T_{10}\to T'_{10}M_8$ decays are
\begin{eqnarray}
A(T_{10}\to T'_{10}M_8)^{TDA}=&&\check{a}H'^{kl}_{mn}(T_{10})^{nij}(T_{10})_{ijk}(M_8)_l^m+\check{b}H'^{kl}_{mn}(T_{10})^{nij}(T_{10})_{ijl}(M_8)_k^m\nonumber\\
&+&\check{c}H'^{kl}_{in}(T_{10})^{nij}(T_{10})_{klj}(M_8)_m^m+\check{d}H'^{kl}_{in}(T_{10})^{nij}(T_{10})_{mkj}(M_8)_l^m\nonumber\\
&+&\check{e}H'^{kl}_{in}(T_{10})^{nij}(T_{10})_{mlj}(M_8)_k^m+\check{f}H'^{kl}_{in}(T_{10})^{nij}(T_{10})_{lkm}(M_8)_j^m\nonumber\\
&+&\check{g}H'^{lk}_{mn}(T_{10})^{nij}(T_{10})_{ilk}(M_8)_j^m+\check{\sigma}H'^{lk}_{ln}(T_{10})^{nij}(T_{10})_{kij}(M_8)_m^m\nonumber\\
&+&\check{\lambda}H'^{lk}_{ln}(T_{10})^{nij}(T_{10})_{mij}(M_8)_k^m+\check{\gamma}H'^{lk}_{ln}(T_{10})^{nij}(T_{10})_{ikm}(M_8)_j^m.
\end{eqnarray}
And  the TDA amplitudes for the $T_{10}\to T'_{10} P_8$ decays are given in Tab. \ref{Tab:T102T10PTDA}.
\begin{table}[h]
\renewcommand\arraystretch{1.42}
\tabcolsep 0.3in
\centering
\caption{The TDA amplitudes of the $T_{10}\to T'_{10} P_8$ decays under the $s\to u\bar{u}d$ transition.  }\vspace{0.1cm}
{\footnotesize
\begin{tabular}{r|ccc}  \hline\hline
 Observables~~~~~~~~~ & Tree& W-exchange &Penguin\\\hline
$\sqrt{6}A(\Omega^-\rightarrow\Xi^{*-}\pi^0)$&$\check{b}$&&$-\check{\beta}$ \\
$\sqrt{3}A(\Omega^-\rightarrow\Xi^{*0}\pi^-)$&$\check{a}$&&$+\check{\beta}$ \\
$\sqrt{6}A(\Omega^-\rightarrow\Sigma^{*0}K^-)$&&$\check{g}$&+$\check{\gamma}$ \\
\hline
$6A(\Xi^{*0}\rightarrow\Sigma^{*0}\pi^0)$&$\check{b}$&$-\check{d}+\check{e}+\check{g}$&$-\check{\beta}+\check{\gamma}$ \\
$3A(\Xi^{*0}\rightarrow\Sigma^{*+}\pi^-)$&$\check{a}$&$+\check{d}$&$+\check{\beta}$ \\
$3A(\Xi^{*0}\rightarrow\Delta^{+}K^-)$&&$\check{f}+\check{g}$&$+\check{\gamma}$ \\
$3A(\Xi^{*0}\rightarrow\Sigma^{*-}\pi^+)$&&$\check{e}$&$+\check{\gamma}$ \\
$3A(\Xi^{*0}\rightarrow\Delta^{0}K^0)$&&$\check{f}$ &$\check{\gamma}$ \\
\hline
$3\sqrt{2}A(\Xi^{*-}\rightarrow\Sigma^{*-}\pi^0)$&$\check{b}$&&$-\check{\beta}-\check{\gamma}$ \\
$3\sqrt{2}A(\Xi^{*-}\rightarrow\Sigma^{*0}\pi^-)$&$\check{a}$&$+\check{g}$&$+\check{\beta}+\check{\gamma}$ \\
$3A(\Xi^{*-}\rightarrow\Delta^{0}K^-)$&&$\check{g}$&$+\check{\gamma}$ \\\hline
$\sqrt{6}A(\Sigma^{*-}\rightarrow\Delta^{-}\pi^0)$&$\check{b}$&&$-\check{\beta}-\check{\gamma}$ \\
$3A(\Sigma^{*-}\rightarrow\Delta^{0}\pi^-)$&$\check{a}$&$+\check{g}$&$+\check{\beta}+\check{\gamma}$ \\\hline
$3\sqrt{2}A(\Sigma^{*+}\rightarrow\Delta^{+}\pi^0)$&$\check{b}$&$-\check{d}+\check{e}+\check{f}+\check{g}$&$-\check{\beta}+\check{\gamma}$ \\
$\sqrt{3}A(\Sigma^{*+}\rightarrow\Delta^{++}\pi^-)$&$\check{a}$&$+\check{d}$&$+\check{\beta}$ \\\hline
$6A(\Sigma^{*0}\rightarrow\Delta^{0}\pi^0)$&$\check{b}$&$-\check{d}+\check{e}-\check{f}+\check{g}$&$-\check{\beta}$ \\
$3\sqrt{2}A(\Sigma^{*0}\rightarrow\Delta^{+}\pi^-)$&$\check{a}$&$+\check{d}+\check{f}+\check{g}$&$+\check{\beta}+\check{\gamma}$ \\\hline
\end{tabular}\label{Tab:T102T10PTDA}}\vspace{1cm}
\end{table}

\begin{table}[t]
\renewcommand\arraystretch{1.42}
\tabcolsep 0.3in
\centering
\caption{The SU(3) IRA amplitudes of the $T_{10}\to T'_{10} P_8$ decays under the $s\to u\bar{u}d$ transition.
 Noted that $\ddot{a}'$ and $\ddot{b}'$ denote $\ddot{a}$ and $\ddot{b}$ terms in $H(4)^{22}_{2}$, respectively.
 }\vspace{0.1cm}
{\footnotesize
\begin{tabular}{r|ccc}  \hline\hline
 Observables &$H(4)_1^{12}=\frac{1}{3}$ & $H(4)_2^{22}=-\frac{1}{3}$ &$H(\bar{2})^{2}=1$\\\hline
$\sqrt{6}A(\Omega^-\rightarrow\Xi^{*-}\pi^0)$&$\ddot{a}$&$-\ddot{a}'$&$-\ddot{e}$ \\
$\sqrt{3}A(\Omega^-\rightarrow\Xi^{*0}\pi^-)$&$\ddot{a}$&&$+\ddot{e}$ \\
$\sqrt{6}A(\Omega^-\rightarrow\Sigma^{*0}K^-)$&$2\ddot{b}$&&$+\ddot{f}$ \\
\hline
$6A(\Xi^{*0}\rightarrow\Sigma^{*0}\pi^0)$&$\ddot{a}+2\ddot{b}$&$-\ddot{a}'$&$-\ddot{e}+\ddot{f}$ \\
$3A(\Xi^{*0}\rightarrow\Sigma^{*+}\pi^-)$&$\ddot{a}+\ddot{c}$&$$&$+\ddot{e}$ \\
$3A(\Xi^{*0}\rightarrow\Delta^{+}K^-)$&$2\ddot{b}+2\ddot{d}$&&$+\ddot{f}$ \\
$3A(\Xi^{*0}\rightarrow\Sigma^{*-}\pi^+)$&$\ddot{c}$&$+\ddot{b}'$&$+\ddot{f}$ \\
$3A(\Xi^{*0}\rightarrow\Delta^{0}K^0)$&$2\ddot{d}$&$+\ddot{b}'$&$\ddot{f}$ \\
\hline
$3\sqrt{2}A(\Xi^{*-}\rightarrow\Sigma^{*-}\pi^0)$&$\ddot{a}$&$-\ddot{a}'-\ddot{b}'$&$-\ddot{e}-\ddot{f}$ \\
$3\sqrt{2}A(\Xi^{*-}\rightarrow\Sigma^{*0}\pi^-)$&$\ddot{a}+2\ddot{b}$&&$+\ddot{e}+\ddot{f}$ \\
$3A(\Xi^{*-}\rightarrow\Delta^{0}K^-)$&$2\ddot{b}$&&$+\ddot{f}$ \\\hline
$\sqrt{6}A(\Sigma^{*-}\rightarrow\Delta^{-}\pi^0)$&$\ddot{a}$&$-\ddot{a}'-\ddot{b}'$&$-\ddot{e}-\ddot{f}$ \\
$3A(\Sigma^{*-}\rightarrow\Delta^{0}\pi^-)$&$\ddot{a}+2\ddot{b}$&&$+\ddot{e}+\ddot{f}$ \\\hline
$3\sqrt{2}A(\Sigma^{*+}\rightarrow\Delta^{+}\pi^0)$&$\ddot{a}+2\ddot{b}+2\ddot{d}$&$-\ddot{a}'$&$-\ddot{e}+\ddot{f}$\\
$\sqrt{3}A(\Sigma^{*+}\rightarrow\Delta^{++}\pi^-)$&$\ddot{a}+\ddot{c}$&&$+\ddot{e}$ \\\hline
$6A(\Sigma^{*0}\rightarrow\Delta^{0}\pi^0)$&$\ddot{a}+2\ddot{b}-2\ddot{d}$&$-\ddot{a}'-\ddot{b}'$&$-\ddot{e}$\\
$3\sqrt{2}A(\Sigma^{*0}\rightarrow\Delta^{+}\pi^-)$&$\ddot{a}+2\ddot{b}+\ddot{c}+2\ddot{d}$&&$+\ddot{e}+\ddot{f}$\\\hline
\end{tabular}\label{Tab:T102T10PIRA}}
\end{table}

Since the SU(3)  IRA amplitudes of the $T_{10}\to T'_{10} M_8$ decays have not been calculated in Ref. \cite{Wang:2019alu}, we give them in this work.  The IRA amplitudes of the $T_{10}\to T'_{10} M_8$ decays are
\begin{eqnarray}
A(T_{10}\to T'_{10}M_8)^{IRA}=&&\ddot{a}H(4)^{lk}_{m}(T_{10})^{ijn}(T_{10})_{ijk}(M_8)_l^m+\ddot{b}H(4)^{lk}_{m}(T_{10})^{ijn}(T_{10})_{ilk}(M_8)_j^m\nonumber\\
&+&\ddot{c}H(4)^{lk}_{j}(T_{10})^{ijn}(T_{10})_{imk}(M_8)_l^m+\ddot{d}H(4)^{lk}_{j}(T_{10})^{ijn}(T_{10})_{mlk}(M_8)_i^m\nonumber\\
&+&\ddot{e}H(\bar{2})^k(T_{10})^{ijn}(T_{10})_{ijm}(M_8)_k^m+\ddot{f}H(\bar{2})^k(T_{10})^{ijn}(T_{10})_{kim}(M_8)_j^m,
\end{eqnarray}
where the non-zore current-current operators in SU(2) flavor space are $H(4)^{12}_{1}=H(4)^{21}_{1}=\frac{1}{3}$ and $H(\bar{2})^2=1$, small penguin contributions of $H(4)^{22}_2=-\frac{1}{3}$ are ignored in our analysis.
And the concrete amplitudes for the $T_{10}\to T'_{10} P_8$ decays are given  in Tab. \ref{Tab:T102T10PIRA}.

Comparing the TDA amplitudes in Tab. \ref{Tab:T102T10PTDA} and the IRA amplitudes in Tab. \ref{Tab:T102T10PIRA}, we find that they  match consistently in the $T_{10}\to T'_{10} M_8$ decays, and the
relations between the two sets of amplitudes are
\begin{eqnarray}
\ddot{a}+\ddot{e}&=&\check{a}+\check{\beta},~~~~~~~~~~~~~2\ddot{a}-\ddot{a}'~=~\check{a}+\check{b},~~~~~~~~~~~~~2\ddot{b}+\ddot{f}~=~\check{g},\nonumber\\
\ddot{c}&=&\check{d}=\check{e},~~~~~~~~~~~~~~~~~~~~~\ddot{d}~=~\frac{1}{2}\check{f},\nonumber\\
\ddot{f}&=&-\ddot{b}',~~~~~~~~~~~~~~~~~~~~~~~~\check{\gamma}~=~0.
\end{eqnarray}

From the Tab.  \ref{Tab:T102T10PTDA}, we have the following amplitude relations
\begin{eqnarray}
A(\Omega^-\rightarrow\Xi^{*-}\pi^0)=\sqrt{3}A(\Xi^{*-}\rightarrow\Sigma^{*-}\pi^0)&=&\sqrt{6}A(\Sigma^{*-}\rightarrow\Delta^-\pi^0),\nonumber\\
\sqrt{2}A(\Omega^-\rightarrow\Sigma^{*0}K^-)=\sqrt{3}A(\Xi^{*-}\rightarrow\Delta^{0}K^-),&&\nonumber\\
\sqrt{6}A(\Xi^{*-}\rightarrow\Sigma^{*0}\pi^-)=\sqrt{3}A(\Sigma^{*-}\rightarrow\Delta^0\pi^-)&=&A(\Omega^-\rightarrow\Xi^{*0}\pi^-)+\sqrt{2}A(\Omega^-\rightarrow\Sigma^{*0}K^-)\nonumber\\
&=&A(\Omega^-\rightarrow\Xi^{*0}\pi^-)+\sqrt{2}A(\Xi^{*-}\rightarrow\Delta^{0}K^-),
\end{eqnarray}
where we have used $\check{\gamma}=0$ and $-\check{d}+\check{e}=0$.

The W-exchange  contributions in $T_{10}\to T'_{10}M_8$  decays are strongly suppressed since the contraction of the flavor antisymmetric current-current operator with a flavor symmetric final state configuration is zero by K\"{o}rner, Pati, Woo (KPW) theorem  \cite{Korner:1970xq,Pati:1970fg}. If we also ignore  the W-exchange contributions in $T_{10}\to T'_{10}M_8$  decays, we can obtain more simple relations between the amplitudes from Tab. \ref{Tab:T102T10PTDA}.
Ignoring  the W-exchange contributions and  in terms of the central values of the  input parameters,  one may get the ratios of their branching ratios, which are listed in the second and forth columns of Tab. \ref{Tab:T102T10PRBr}. In the second and forth columns of Tab. \ref{Tab:T102T10PRBr}, we take $\mathcal{B}(\Omega^-\rightarrow\Xi^{*-}\pi^0)$ and $\mathcal{B}(\Omega^-\rightarrow\Xi^{*0}\pi^-)$ as a unit for $\mathcal{B}(T_{10}\to T'_{10}\pi^0)$ and $\mathcal{B}(T_{10}\to T'_{10}\pi^-)$, respectively.

Using the experimental upper limit $\mathcal{B}(\Omega^-\rightarrow\Xi^{*0}\pi^-)<7\times10^{-5}$ at 90\% CL \cite{PDG2018} and  considering the $1\sigma$ error of relevant input parameters,  we may predict the upper limits of  $\mathcal{B}(T_{10}\to T'_{10}\pi^-)$, which are listed in the last columns of Tab. \ref{Tab:T102T10PRBr}.  We can see that all branching ratios are very tiny and they can not be measured in recent experiments such as BESIII and LHCb.

\begin{table}[t]
\renewcommand\arraystretch{1.56}
\tabcolsep 0.2in
\centering
\caption{ Branching ratios of the $T_{10}\to T'_{10} P_8$ decays under the $s\to u\bar{u}d$ transition. The ratios of the branching ratios are obtained in terms of the central values of the  input parameters, and the upper limits of $\mathcal{B}(T_{10}\to T'_{10}\pi^-)$ are obtained with the $1\sigma$ error of input parameters.}\vspace{0.1cm}
{\footnotesize
\begin{tabular}{rc||ccc}  \hline\hline
 Decay modes &  $\frac{\mathcal{B}(T_{10}\to T'_{10}\pi^0)}{\mathcal{B}(\Omega^-\rightarrow\Xi^{*-}\pi^0)}$ &Decay modes& $\frac{\mathcal{B}(T_{10}\to T'_{10}\pi^-)}{\mathcal{B}(\Omega^-\rightarrow\Xi^{*0}\pi^-)}$  & Upper Limits of $\mathcal{B}(T_{10}\to T'_{10}\pi^-)$ \\\hline
$\Omega^-\rightarrow\Xi^{*-}\pi^0$&$1$&$\Omega^-\rightarrow\Xi^{*0}\pi^-$&$1$&$7\times10^{-5}$ \\
\hline
$\Xi^{*0}\rightarrow\Sigma^{*0}\pi^0$&$1.89\times10^{-12}$&$\Xi^{*0}\rightarrow\Sigma^{*+}\pi^-$&$7.87\times10^{-12}$&$2.13\times10^{-15}$ \\
\hline
$\Xi^{*-}\rightarrow\Sigma^{*-}\pi^0$&$3.35\times10^{-12}$&$\Xi^{*-}\rightarrow\Sigma^{*0}\pi^-$&$5.08\times10^{-12}$&$1.47\times10^{-15}$ \\
\hline
$\Sigma^{*-}\rightarrow\Delta^{-}\pi^0$&$5.16\times10^{-12}$&$\Sigma^{*-}\rightarrow\Delta^{0}\pi^-$&$3.98\times10^{-12}$&$1.15\times10^{-15}$\\\hline
$\Sigma^{*+}\rightarrow\Delta^{+}\pi^0$&$1.27\times10^{-12}$&$\Sigma^{*+}\rightarrow\Delta^{++}\pi^-$&$7.78\times10^{-12}$&$2.27\times10^{-15}$ \\\hline
$\Sigma^{*0}\rightarrow\Delta^{0}\pi^0$&$6.95\times10^{-13}$&$\Sigma^{*0}\rightarrow\Delta^{+}\pi^-$&$1.46\times10^{-12}$&$4.57\times10^{-16}$\\\hline
\end{tabular}\label{Tab:T102T10PRBr}}
\end{table}

\subsection{$T_{8}\to T'_{8}M_8$  weak decays}
Topological diagrams for $T_{8}\to T'_{8}M_8$  are also displayed in Fig. \ref{fig:NLTP}. Since both initial state $T_8$ and final state $T'_8$ are all not fully symmetric or antisymmetric in flavor space, there are many amplitudes corresponding to one topological diagram, as a result, 92 amplitudes for the $T_{8}\to T'_{8}M_8$ decays correspond to 10 topological diagrams in Fig. \ref{fig:NLTP}.   The TDA amplitudes of the $T_{8}\to T'_{8}M_8$ decays can be constructed as
{\small
\begin{eqnarray}
 A(T_{8}\to T'_8M_8)^{TDA}&=&\tilde{a}_{1}H^{kl}_{mn}(T_{8})^{[ij]n}(T_{8})_{[ij]k}(M_8)_l^m+\tilde{a}_{2}H^{kl}_{mn}(T_{8})^{[ij]n}(T_{8})_{[ik]j}(M_8)_l^m
\nonumber\\
&+&\tilde{a}_{3}H^{kl}_{mn}(T_{8})^{[in]j}(T_{8})_{[ij]k}(M_8)_l^m+\tilde{a}_{4}H^{kl}_{mn}(T_{8})^{[in]j}(T_{8})_{[ik]j}(M_8)_l^m+\tilde{a}_{5}H^{kl}_{mn}(T_{8})^{[in]j}(T_{8})_{[kj]i}(M_8)_l^m\nonumber\\
&+&\tilde{b}_{1}H^{kl}_{mn}(T_{8})^{[ij]n}(T_{8})_{[ij]l}(M_8)_k^m+\tilde{b}_{2}H^{kl}_{mn}(T_{8})^{[ij]n}(T_{8})_{[il]j}(M_8)_k^m\nonumber\\
&+&\tilde{b}_{3}H^{kl}_{mn}(T_{8})^{[in]j}(T_{8})_{[ij]l}(M_8)_k^m+\tilde{b}_{4}H^{kl}_{mn}(T_{8})^{[in]j}(T_{8})_{[il]j}(M_8)_k^m+\tilde{b}_{5}H^{kl}_{mn}(T_{8})^{[in]j}(T_{8})_{[lj]i}(M_8)_k^m\nonumber\\
&+&\tilde{c}_{1}H^{lk}_{in}(T_{8})^{[ij]n}(T_{8})_{[jk]l}(M_8)_m^m+\tilde{c}_{2}H^{lk}_{in}(T_{8})^{[ij]n}(T_{8})_{[jl]k}(M_8)_m^m+\tilde{c}_{3}H^{lk}_{in}(T_{8})^{[ij]n}(T_{8})_{[kl]j}(M_8)_m^m\nonumber\\
&+&\tilde{c}_{4}H^{lk}_{in}(T_{8})^{[in]j}(T_{8})_{[jk]l}(M_8)_m^m+\tilde{c}_{5}H^{lk}_{in}(T_{8})^{[in]j}(T_{8})_{[jl]k}(M_8)_m^m+\tilde{c}_{6}H^{lk}_{in}(T_{8})^{[in]j}(T_{8})_{[kl]j}(M_8)_m^m\nonumber\\
&+&\tilde{c}_{7}H^{lk}_{in}(T_{8})^{[jn]i}(T_{8})_{[jk]l}(M_8)_m^m+\tilde{c}_{8}H^{lk}_{in}(T_{8})^{[jn]i}(T_{8})_{[jl]k}(M_8)_m^m+\tilde{c}_{9}H^{lk}_{in}(T_{8})^{[jn]i}(T_{8})_{[kl]j}(M_8)_m^m\nonumber\\
&+&\tilde{d}_{1}H^{lk}_{in}(T_{8})^{[ij]n}(T_{8})_{[jm]k}(M_8)_l^m+\tilde{d}_{2}H^{lk}_{in}(T_{8})^{[ij]n}(T_{8})_{[jk]m}(M_8)_l^m+\tilde{d}_{3}H^{lk}_{in}(T_{8})^{[ij]n}(T_{8})_{[km]j}(M_8)_l^m\nonumber\\
&+&\tilde{d}_{4}H^{lk}_{in}(T_{8})^{[in]j}(T_{8})_{[jm]k}(M_8)_l^m+\tilde{d}_{5}H^{lk}_{in}(T_{8})^{[in]j}(T_{8})_{[jk]m}(M_8)_l^m+\tilde{d}_{6}H^{lk}_{in}(T_{8})^{[in]j}(T_{8})_{[km]j}(M_8)_l^m\nonumber\\
&+&\tilde{d}_{7}H^{lk}_{in}(T_{8})^{[jn]i}(T_{8})_{[jm]k}(M_8)_l^m+\tilde{d}_{8}H^{lk}_{in}(T_{8})^{[jn]i}(T_{8})_{[jk]m}(M_8)_l^m+\tilde{d}_{9}H^{lk}_{in}(T_{8})^{[jn]i}(T_{8})_{[km]j}(M_8)_l^m\nonumber\\
&+&\tilde{e}_{1}H^{lk}_{in}(T_{8})^{[ij]n}(T_{8})_{[jm]l}(M_8)_k^m+\tilde{e}_{2}H^{lk}_{in}(T_{8})^{[ij]n}(T_{8})_{[jl]m}(M_8)_k^m+\tilde{e}_{3}H^{lk}_{in}(T_{8})^{[ij]n}(T_{8})_{[lm]j}(M_8)_k^m\nonumber\\
&+&\tilde{e}_{4}H^{lk}_{in}(T_{8})^{[in]j}(T_{8})_{[jm]l}(M_8)_k^m+\tilde{e}_{5}H^{lk}_{in}(T_{8})^{[in]j}(T_{8})_{[jl]m}(M_8)_k^m+\tilde{e}_{6}H^{lk}_{in}(T_{8})^{[in]j}(T_{8})_{[lm]j}(M_8)_k^m\nonumber\\
&+&\tilde{e}_{7}H^{lk}_{in}(T_{8})^{[jn]i}(T_{8})_{[jm]l}(M_8)_k^m+\tilde{e}_{8}H^{lk}_{in}(T_{8})^{[jn]i}(T_{8})_{[jl]m}(M_8)_k^m+\tilde{e}_{9}H^{lk}_{in}(T_{8})^{[jn]i}(T_{8})_{[lm]j}(M_8)_k^m\nonumber\\
&+&\tilde{f}_{1}H^{lk}_{in}(T_{8})^{[ij]n}(T_{8})_{[kl]m}(M_8)_j^m+\tilde{f}_{2}H^{lk}_{in}(T_{8})^{[ij]n}(T_{8})_{[km]l}(M_8)_j^m+\tilde{f}_{3}H^{lk}_{in}(T_{8})^{[ij]n}(T_{8})_{[ml]k}(M_8)_j^m\nonumber\\
&+&\tilde{f}_{4}H^{lk}_{in}(T_{8})^{[in]j}(T_{8})_{[kl]m}(M_8)_j^m+\tilde{f}_{5}H^{lk}_{in}(T_{8})^{[in]j}(T_{8})_{[km]l}(M_8)_j^m+\tilde{f}_{6}H^{lk}_{in}(T_{8})^{[in]j}(T_{8})_{[ml]k}(M_8)_j^m\nonumber\\
&+&\tilde{f}_{7}H^{lk}_{in}(T_{8})^{[jn]i}(T_{8})_{[kl]m}(M_8)_j^m+\tilde{f}_{8}H^{lk}_{in}(T_{8})^{[jn]i}(T_{8})_{[km]l}(M_8)_j^m+\tilde{f}_{9}H^{lk}_{in}(T_{8})^{[jn]i}(T_{8})_{[ml]k}(M_8)_j^m\nonumber\\
&+&\tilde{g}_{1}H^{lk}_{mn}(T_{8})^{[ij]n}(T_{8})_{[ik]l}(M_8)_j^m+\tilde{g}_{2}H^{lk}_{mn}(T_{8})^{[ij]n}(T_{8})_{[il]k}(M_8)_j^m+\tilde{g}_{3}H^{lk}_{mn}(T_{8})^{[ij]n}(T_{8})_{[lk]i}(M_8)_j^m\nonumber\\
&+&\tilde{g}_{4}H^{lk}_{mn}(T_{8})^{[in]j}(T_{8})_{[ik]l}(M_8)_j^m+\tilde{g}_{5}H^{lk}_{mn}(T_{8})^{[in]j}(T_{8})_{[il]k}(M_8)_j^m+\tilde{g}_{6}H^{lk}_{mn}(T_{8})^{[in]j}(T_{8})_{[lk]i}(M_8)_j^m\nonumber\\
&+&\tilde{g}_{7}H^{lk}_{mn}(T_{8})^{[jn]i}(T_{8})_{[ik]l}(M_8)_j^m+\tilde{g}_{8}H^{lk}_{mn}(T_{8})^{[jn]i}(T_{8})_{[il]k}(M_8)_j^m+\tilde{g}_{9}H^{lk}_{mn}(T_{8})^{[jn]i}(T_{8})_{[lk]i}(M_8)_j^m\nonumber\\
&+&\tilde{\sigma}_{1}H^{lk}_{ln}(T_{8})^{[ij]n}(T_{8})_{[ij]k}(M_8)_m^m+\tilde{\sigma}_{2}H^{lk}_{ln}(T_{8})^{[ij]n}(T_{8})_{[ik]j}(M_8)_m^m\nonumber\\
&+&\tilde{\sigma}_{3}H^{lk}_{ln}(T_{8})^{[in]j}(T_{8})_{[ij]k}(M_8)_m^m+\tilde{\sigma}_{4}H^{lk}_{ln}(T_{8})^{[in]j}(T_{8})_{[ik]j}(M_8)_m^m+\tilde{\sigma}_{5}H^{lk}_{ln}(T_{8})^{[in]j}(T_{8})_{[kj]i}(M_8)_m^m\nonumber\\
&+&\tilde{\lambda}_{1}H^{lk}_{ln}(T_{8})^{[ij]n}(T_{8})_{[ij]m}(M_8)_k^m+\tilde{\lambda}_{2}H^{lk}_{ln}(T_{8})^{[ij]n}(T_{8})_{[im]j}(M_8)_k^m\nonumber\\
&+&\tilde{\lambda}_{3}H^{lk}_{ln}(T_{8})^{[in]j}(T_{8})_{[ij]m}(M_8)_k^m+\tilde{\lambda}_{4}H^{lk}_{ln}(T_{8})^{[in]j}(T_{8})_{[im]j}(M_8)_k^m+\tilde{\lambda}_{5}H^{lk}_{ln}(T_{8})^{[in]j}(T_{8})_{[mj]i}(M_8)_k^m\nonumber\\
&+&\tilde{\gamma}_{1}H^{lk}_{in}(T_{8})^{[ij]n}(T_{8})_{[ik]m}(M_8)_j^m+\tilde{\gamma}_{2}H^{lk}_{in}(T_{8})^{[ij]n}(T_{8})_{[im]k}(M_8)_j^m+\tilde{\gamma}_{3}H^{lk}_{in}(T_{8})^{[ij]n}(T_{8})_{[mk]i}(M_8)_j^m\nonumber\\
&+&\tilde{\gamma}_{4}H^{lk}_{in}(T_{8})^{[in]j}(T_{8})_{[ik]m}(M_8)_j^m+\tilde{\gamma}_{5}H^{lk}_{in}(T_{8})^{[in]j}(T_{8})_{[im]k}(M_8)_j^m+\tilde{\gamma}_{6}H^{lk}_{in}(T_{8})^{[in]j}(T_{8})_{[mk]i}(M_8)_j^m\nonumber\\
&+&\tilde{\gamma}_{7}H^{lk}_{in}(T_{8})^{[jn]i}(T_{8})_{[ik]m}(M_8)_j^m+\tilde{\gamma}_{8}H^{lk}_{in}(T_{8})^{[jn]i}(T_{8})_{[im]k}(M_8)_j^m+\tilde{\gamma}_{9}H^{lk}_{in}(T_{8})^{[jn]i}(T_{8})_{[mk]i}(M_8)_j^m.
\label{Eq:T82T8M8TDA}
\end{eqnarray}}
From  Eq. (\ref{Eq:T82T8M8TDA}), we can get the TDA amplitudes of different decay modes.  Since many parameters are not independent,  we give the redefinitions
\begin{eqnarray}
X_1&\equiv&2\tilde{x}_1+2\tilde{x}_2+2\tilde{x}_3+\tilde{x}_4+\tilde{x}_5, \nonumber\\
X_2&\equiv&\tilde{x}_4-\tilde{x}_5,
\end{eqnarray}
where $X=A,B,\Sigma,\Lambda$ corresponding to $\tilde{a}_i,\tilde{b}_i,\tilde{\sigma}_i,\tilde{\lambda}_i$, respectively,  and
\begin{eqnarray}
Y_1&\equiv&\tilde{y}_1+\tilde{y}_3+\tilde{y}_4+\tilde{y}_6,\nonumber\\
Y_2&\equiv&\tilde{y}_4+\tilde{y}_6+\tilde{y}_7+\tilde{y}_9,\nonumber\\
Y_3&\equiv&\tilde{y}_4+\tilde{y}_5+\tilde{y}_7+\tilde{y}_8,\nonumber\\
Y_4&\equiv&\tilde{y}_2-\tilde{y}_3+\tilde{y}_5-\tilde{y}_6,
\end{eqnarray}
with $Y=C,D,E,F,G,\Gamma$ for $\tilde{c}_i,\tilde{d}_i,\tilde{e}_i,\tilde{f}_i,\tilde{g}_i,\tilde{\gamma}_i$, respectively. And then the parameters related to $T_{8}\to T'_{8}M_8$ decays studied in this paper are reduced to 26. The corresponding amplitudes are listed in Tab. \ref{Tab:T82T8PTDA}.
\begin{table}[b]
\renewcommand\arraystretch{1.56}
\tabcolsep 0.1in
\centering
\caption{The TDA amplitudes of the $T_8\to T'_8 P_8$ decays under the $s\to u\bar{u}d$ transition.}\vspace{0.1cm}
{\footnotesize
\begin{tabular}{r|ccc|c}  \hline\hline
 Amplitudes~~~~~ & Tree & W-exchange&Penguin& Ratios of amplitude moduli   \\\hline
$\sqrt{2}A(\Sigma^+\rightarrow p\pi^0)$&$-B_2$&$+D_2+E_2-E_3-F_2-G_2$&$+2\Lambda_2-\Gamma_2$&2.02\\
$A(\Sigma^+\rightarrow n\pi^+)$&&$-E_3-F_1$&$-\Gamma_3$&1.01 \\
$A(\Sigma^-\rightarrow n\pi^-)$&$-A_2$&$+G_2-G_3$&$-2\Lambda_2+\Gamma_2-\Gamma_3$ &1.00\\
$\sqrt{2}A(\Sigma^0\rightarrow p\pi^-)$&$A_2$&$+D_3-F_2-G_2$&$+2\Lambda_2-\Gamma_2$&2.02 \\
$2A(\Sigma^0\rightarrow n\pi^0)$&$-B_2$&$+D_2-D_3+E_2+F_1-G_2+G_3$&$+2\Lambda_2-\Gamma_2+2\Gamma_3$&2.83\\\hline
$\sqrt{6}A(\Lambda^0\rightarrow p\pi^-)$&$-A_1$&$^{+2D_1-D_3+2D_4}_{-2F_1+F_2-2G_1+G_2}$&$-2\Lambda_1-2\Gamma_1+\Gamma_2$&9.95 \\
$2\sqrt{3}A(\Lambda^0\rightarrow n\pi^0)$&$-B_1$&$^{-D_2+D_3-2D_4+2E_1-E_2}_{+2F_1-F_3+2F_4-G_2+G_3-2G_4}$&$+2\Lambda_1+2\Gamma_1-\Gamma_2$&14.17\\\hline
$\sqrt{6}A(\Xi^-\rightarrow \Lambda^0\pi^-)$&$\frac{1}{2}A_1+\frac{3}{2}A_2$&$+G_1-2G_2+G_3-G_4$& $^{+(\Lambda_1+3\Lambda_2)}_{+(\Gamma_1-2\Gamma_2+\Gamma_3-\Gamma_4)}$&9.66\\
$2\sqrt{3}A(\Xi^0\rightarrow \Lambda^0\pi^0)$&$\frac{1}{2}B_1+\frac{3}{2}B_2$&$^{-D_1+D_4-E_1+E_4}_{-G_1+2G_2-G_3+G_4}$ &$^{-(\Lambda_1+3\Lambda_2)}_{-(\Gamma_1-2\Gamma_2+\Gamma_3-\Gamma_4)}$&15.24\\\hline
\end{tabular}}\label{Tab:T82T8PTDA}
\end{table}
Even if neglecting 6 small penguin amplitudes, there still are 20 parameters.  Since the parameters are  too much,  it's difficult for us to find many relations between different decay amplitudes.  And there is  only one relation for the decay amplitudes
\begin{eqnarray}
\sqrt{2}A(\Sigma^+\rightarrow p\pi^0)=A(\Sigma^+\rightarrow n\pi^+)+A(\Sigma^-\rightarrow n\pi^-)+\sqrt{2}A(\Sigma^0\rightarrow p\pi^-)+2A(\Sigma^0\rightarrow n\pi^0),
\end{eqnarray}
in which includes the amplitudes of the tree diagrams, the W-exchange diagrams and the penguin diagrams.

Many $T_{8}\to T'_{8}P_8$ decays have been measured \cite{PDG2018}
\begin{eqnarray}
\mathcal{B}(\Sigma^+\rightarrow p\pi^0)&=&(51.57\pm0.30)\times10^{-2}, \nonumber\\
\mathcal{B}(\Sigma^+\rightarrow n\pi^+)&=&(48.31\pm0.30)\times10^{-2},\nonumber\\
\mathcal{B}(\Sigma^-\rightarrow n\pi^-)&=&(99.848\pm0.005)\times10^{-2},\nonumber\\
\mathcal{B}(\Lambda^0\rightarrow p\pi^-)&=&(63.9\pm0.5)\times10^{-2}, \nonumber\\
\mathcal{B}(\Lambda^0\rightarrow n\pi^0)&=&(35.8\pm0.5)\times10^{-2},\nonumber\\
\mathcal{B}(\Xi^-\rightarrow \Lambda^0\pi^-)&=&(99.887\pm0.035)\times10^{-2},\nonumber\\
\mathcal{B}(\Xi^0\rightarrow \Lambda^0\pi^0)&=&(99.524\pm0.012)\times10^{-2}.\label{Eq:T8tT8pP8Exp}
\end{eqnarray}
For not-yet-measured $\Sigma^0\to p\pi^-,n\pi^0$ decays,   we may get the branching ratios in terms of the isospin relations \cite{Wang:2019alu}
\begin{eqnarray}
\mathcal{B}(\Sigma^0\rightarrow p\pi^-)&=&(4.82\pm0.50)\times10^{-10}, \nonumber\\
\mathcal{B}(\Sigma^0\rightarrow n\pi^0)&=&(2.41\pm0.26)\times10^{-10}. \label{Eq:T8tT8pP8IR}
\end{eqnarray}
These measured branching ratios and obtained branching ratios  from the isospin relations will help us to understand decay amplitudes of the tree diagrams and W-exchange diagrams.  Using the central values in Eqs. (\ref{Eq:T8tT8pP8Exp}-\ref{Eq:T8tT8pP8IR}) and the relation between the branching ratios and the moduli of the decay amplitudes in Eq. (\ref{Eq:Br}), one can backward obtain the ratios between  the different moduli of decay amplitudes, which are given
in the last column of Tab. \ref{Tab:T82T8PTDA}. Noted that the ratios are obtained  by using $|A(\Sigma^-\rightarrow n\pi^-)|$ as a unit, in which we including the coefficients in front of the amplitudes in the first column of  Tab. \ref{Tab:T82T8PTDA}. For an example, in the second line and the last column of   Tab. \ref{Tab:T82T8PTDA}, the ratio 2.02 is obtained by using $\frac{\sqrt{2}|A(\Sigma^+\rightarrow p\pi^0)|}{|A(\Sigma^-\rightarrow n\pi^-)|}$, which is also equal to $\frac{|-B_2+D_2+E_2-E_3-F_2-G_2+2\Lambda_2-\Gamma_2|}{|-A_2+G_2-G_3-2\Lambda_2+\Gamma_2-\Gamma_3|}$.  After ignoring the penguin contributions since they are small, we have the following remarks for Tab. \ref{Tab:T82T8PTDA}.
\begin{itemize}

\item As shown in the third line, there is no tree diagram contribution to the $\Sigma^+\rightarrow n\pi^+$ decay, nevertheless $|A(\Sigma^+\rightarrow n\pi^+)|$ may compare with others.  So the W-exchange contributions  can be as large as some tree diagram contributions.

\item For the tree diagram amplitudes shown in Fig. \ref{fig:NLTP} (a-b), $A_i \propto C_2+C_1/N_C\approx1.153$ and $B_i \propto C_1+C_2/N_C\approx-0.171$, $i.e.$, $B_i$ are suppressed by the color-flavor factor.
Comparing $\sqrt{6}|A(\Lambda^0\rightarrow p\pi^-)|$ with $2\sqrt{3}|A(\Lambda^0\rightarrow n\pi^0)|$ or comparing $\sqrt{6}|A(\Xi^-\rightarrow \Lambda^0\pi^-)|$ with $2\sqrt{3}|A(\Xi^0\rightarrow \Lambda^0\pi^0)|$, the formers included $A_i$   are smaller than the latter  included $B_i$,  so the W-exchange diagrams  give large and even dominant contribution to relevant decay amplitudes.

\end{itemize}

\section{SUMMARY}
In this work, we have  analyzed the two-body nonleptonic decays of light baryon octet and decuplet
by using the TDA  under the SU(3) flavor symmetry. Comparing with the IRA, the TDA is more intuitive and gives a better understanding of dynamics.

We have found that the TDA and the IRA   can  match  in the $T_{10}\to T^{(')}_{8,10}P_8$ weak decays.
For the $T_{10}\to T^{(')}_{8,10}P_8$  weak decays, the  TPA gives more simple  predictions than the IRA since the W-exchange contributions are strongly suppressed. We have predicted all relevant decay branching ratios  of the $T_{10}\to T_8P_8$ by using the TDA and also with the help of the IRA. For the $T_{10}\to T'_{10}P_8$ decays,  we have obtained some amplitude relations and  the ratios of the branching ratios  by using the TDA and the  IRA,  and then  gotten   the five upper limits of  $\mathcal{B}(T_{10}\to T'_{10}\pi^-)$  by using the experimental upper limit of $\mathcal{B}(\Omega^-\to \Xi^{*0}\pi^-)$.  Nevertheless,  since the lifetimes of $\Xi^{*0,-}$ and $\Sigma^{*0,\pm}$ are very small, all new branching ratio predictions, which are on the order of $\mathcal{O}(10^{-16}-10^{-12})$,  are too small to be searched in current experiments such as the BESIII and LHCb.

For the $T_{8}\to T'_{8}P_8$ decays, since both initial states $T_8$ and final states $T'_8$ are all not fully symmetric or antisymmetric in flavor space, these decays are quite complex in the theory, and the  TPA gives  more complex  predictions than the IRA.
Using some  experiential  branching ratios and other branching ratios  obtained from the isospin relations, we analyze the contribution size of the tree diagrams and the W-exchange diagrams, and we have  found  that the W-exchange diagrams  give large and even dominant contributions to some decays.

\section*{ACKNOWLEDGEMENTS}
The work was supported by the National Natural Science Foundation of China (Contract No. 11675137) and the Key Scientific Research Projects of Colleges and Universities in Henan Province (Contract No. 18A140029).

\section*{References}

\renewcommand{\baselinestretch}{1.45}

\end{document}